\documentclass[prd,amsmath,amssymb,floatfix,superscriptaddress,notitlepage,nofootinbib,preprintnumbers]{revtex4-1}
\usepackage{amsfonts,amssymb,amsmath,graphicx,color}
\begin{document}
\newcommand{\f}[2]{\frac{#1}{#2}}
\newcommand{\be}{\begin{equation}}
\newcommand{\ee}{\end{equation}}
\newcommand{\Mpl}{M_{\rm Pl}}
\newcommand{\E}{\mathcal{E}}
\renewcommand{\L}{\mathcal{L}}
\newcommand{\EC}{\E_C}
\newcommand{\ED}{\E_D}
\newcommand{\ELC}{\E_{C,\Lambda}}
\newcommand{\ELD}{\E_{D,\Lambda}}
\newcommand{\pa}{\partial}

\title{
Black hole solutions in shift-symmetric degenerate higher-order scalar-tensor theories
}

\author{Masato Minamitsuji}
\affiliation{Centro de Astrof\'{\i}sica e Gravita\c c\~ao  - CENTRA, Departamento de F\'{\i}sica, Instituto Superior T\'ecnico - IST,
Universidade de Lisboa - UL, Av. Rovisco Pais 1, 1049-001 Lisboa, Portugal}

\author{James Edholm}
\affiliation{
Physics Department, Lancaster University, Lancaster, LA1 4YW
}

\begin{abstract}

We show that 
most classes of shift-symmetric degenerate higher-order scalar-tensor (DHOST) theories which satisfy certain degeneracy conditions are not compatible with the conditions for the existence of exact black hole solutions with a linearly time-dependent scalar field whose canonical kinetic term takes a constant value. Combined with constraints from the propagation speed of gravitational waves, our results imply that cubic DHOST theories are strongly disfavoured and that pure quadratic theories are likely to be the most viable class of DHOST theories.
We find  exact static and spherically symmetric (Schwarzschild and Schwarzschild-(anti-)de Sitter) black hole solutions in all shift-symmetric 
higher-derivative scalar-tensor theories which contain up to cubic order terms of the second-order derivatives of the scalar field,
especially the full class of Horndeski
and Gleyzes-Langlois-Piazza-Vernizzi theories. After deriving the conditions for the coupling functions in the DHOST Lagrangian that allow the exact solutions, we clarify their compatibility with the degeneracy conditions.

\end{abstract}

\maketitle

\section{Introduction}
\label{sec1}


Degenerate Higher-Order Scalar-Tensor (DHOST) theories~\cite{Langlois:2015cwa,Achour:2016rkg,BenAchour:2016fzp,Crisostomi:2017lbg,Langlois:2017dyl} 
provide a general framework of single-field scalar-tensor theories which contain 
higher order terms of second-order derivatives of the scalar field. They allow higher-order Euler-Lagrange equations but possess three degrees of freedom,
i.e. two metric and one scalar degrees of freedom, 
without Ostrogradsky ghosts~\cite{Woodard:2015zca},
as the system of equations reduces to that with the second order derivatives after a suitable field redefinition.
DHOST theories represent the extension of similar analyses in the context of analytical mechanics~\cite{Motohashi:2016ftl,Motohashi:2017eya,Motohashi:2018pxg}
(See Refs. \cite{Langlois:2018dxi,Kobayashi:2019hrl} for reviews).

The direct detection of gravitational waves from binary black hole (BH) and neutron star (NS) mergers~\cite{TheLIGOScientific:2017qsa,GBM:2017lvd,Monitor:2017mdv} means that the
application of higher-derivative scalar-tensor theories to BH solutions is of great interest. 
Depending on the profile of the scalar field, these solutions can be classified into several classes. 
Besides BH solutions in general relativity (GR) with a
constant scalar field, 
higher-derivative scalar-tensor theories also allow BH solutions which are absent in GR.
Stealth Schwarzschild solutions were found in~\cite{Babichev:2013cya,Kobayashi:2014eva} 
for the shift-symmetric (i.e. symmetric under $\phi\to\phi+c$ where $c$ is a constant) subclasses of the Horndeski theory~\cite{Horndeski:1974wa},
which are also subclasses of the DHOST theories,
for the linearly time-dependent scalar field profile $\phi(t,r)=q\hspace{0.4mm}t+\psi(r)$,
with the assumption of a constant kinetic term $X:=g^{\mu\nu}\partial_\mu\phi\partial_\nu\phi$. 
Here, ``stealth'' means that
the metric is independent of 
the scalar field sector of the Lagrangian
and the existence of the scalar field is hidden in the spacetime geometry.
Stealth BH solutions have also been studied in several different contexts 
\cite{AyonBeato:2004ig,Mukohyama:2005rw,Gaete:2013oda,Chagoya:2016aar}.

BH solutions with metrics different from those in GR
were also found in the shift-symmetric theories with the scalar profile with nonconstant $X$.
An asymptotically flat solution
with the static scalar field $\phi=\phi(r)$
was also constructed 
in the Einstein-scalar-Gauss-Bonnet theory with linear coupling,
which is a subclass of the shift-symmetric Horndeski theories~\cite{Sotiriou:2013qea,Sotiriou:2014pfa}.
The key for the existence of this solution
is the violation of one of the assumptions in the no-hair theorem for the shift-symmetric theories~\cite{Hui:2012qt},
that the coupling functions and their derivatives are nonsingular as $X\to 0$.
Ref.~\cite{Babichev:2017guv} presented the extension of the no-hair theorem 
to the shift-symmetric Gleyzes-Langlois-Piazza-Vernizzi (GLPV) theories (also known as beyond-Horndeski theories)~\cite{Gleyzes:2014dya}
and asymptotically flat hairy BH solutions that circumvent the assumptions of the theorem.
In contrast to~\cite{Babichev:2013cya,Kobayashi:2014eva},
hairy BH solutions obtained in Refs. \cite{Sotiriou:2013qea,Sotiriou:2014pfa,Babichev:2017guv}
possess the profiles of the scalar field with nonconstant $X$.


Another interesting solution is the self-tuned Schwarzschild-de Sitter solution,
in which the metric contains an effective cosmological constant 
independent of the bare cosmological constant in the action,
and hence the spacetime metric may not be affected by a sudden change
of the bare cosmological constant via a phase transition~\cite{Charmousis:2011bf,Charmousis:2011ea,Martin-Moruno:2015bda}.
The self-tuned Schwarzschild-de Sitter solutions were found in the shift-symmetric Horndeski~\cite{Babichev:2013cya,Kobayashi:2014eva} and GLPV~\cite{Babichev:2016kdt,Babichev:2017lmw} theories.

In the context of quadratic DHOST theories,
the conditions for the existence of the Schwarzschild and 
Schwarzschild (anti)-de Sitter solutions
under the assumptions of a linearly time-dependent profile of the scalar field 
$\phi(t,r)=q\hspace{0.5mm} t+\psi(r)$ and a constant value of $X=X_0$
were obtained in Refs. \cite{Motohashi:2019sen,BenAchour:2018dap}.
Odd- and even-mode perturbations about the static and spherically BH solutions  
in pure quadratic DHOST theories 
were analyzed in Refs. \cite{Takahashi:2019oxz,deRham:2019gha}.
The analysis of static BH solutions in quadratic DHOST theories
was recently extended to Kerr and Kerr- (anti-) de Sitter solutions \cite{Charmousis:2019vnf}.
In this paper, 
we clarify the conditions for the existence of the Schwarzschild and 
Schwarzschild-(anti-) de Sitter BH solutions 
in more general classes of the DHOST theories including cubic DHOST theories.
Our analysis follows that for the shift-symmetric quadratic DHOST theories in Ref. \cite{Motohashi:2019sen,BenAchour:2018dap,Takahashi:2019oxz}.
We also present the detailed derivation process of these conditions, and give explicit models and solutions.

As shown in Ref. \cite{Langlois:2017dyl}
cubic DHOST theories 
do not appear to satisfy the recent constraint on the propagation speed of gravitational waves $c_g$,
obtained from the coincident measurements
of gravitational waves emitted from a binary NS merger
and its associated short gamma-ray burst
\cite{Monitor:2017mdv,Creminelli:2017sry,Ezquiaga:2017ekz,Baker:2017hug,Sakstein:2017xjx},
indicating 
that the difference between $c_g$ and the speed of light 
is at most ${\cal O}(10^{-15} )$.
However, there are two reasons why the above constraints do not directly apply to our problem and therefore 
it is worthwhile to investigate exact BH solutions in cubic DHOST theories. 
The first 
is that the constraint is based on the assumption that the scalar field has cosmological origins and therefore is distributed over cosmological scales. 
However,  
the scalar field may be localized only in the vicinity of BHs or NSs,
or may not have a cosmological origin.
The second argument is that
the frequency band of LIGO $\sim 100{\rm Hz}$ is very close to the cut-off scale of 
the effective field theory of dark energy \cite{deRham:2018red},
and hence the constraints from LIGO may not be directly applicable to the Horndeski or DHOST theories
if they provide the effective field theory description for dark energy.

The paper is constructed as follows:
In Sec. \ref{sec2}, we review the DHOST theories.
In Sec. \ref{sec3} and Sec. \ref{sec4},
we derive the conditions for the existence of the Schwarzschild and Schwarzschild-(anti-)de Sitter solutions
in the pure cubic and full DHOST theories.
We check the compatibility of the above conditions 
with the degeneracy conditions.
In Sec. \ref{sec5}, we discuss the limits to special subclasses known as Horndeski theory~\cite{Horndeski:1974wa} and GLPV theory~\cite{Gleyzes:2014dya}.

\section{Set-up}
\label{sec2}
\subsection{DHOST theories}

We consider the most general theories which are composed of terms up to cubic order of
the second order covariant derivatives of the scalar field 
\begin{eqnarray}
 \label{qdaction}
\hspace{-1mm}S=
\hspace{-0.6mm}\int\hspace{-0.4mm} d^4x \sqrt{-g} 
\bigg[
   F_0(\phi,X)
 \hspace{-0.4mm}+\hspace{-0.4mm} F_1(\phi,X) \Box\phi 
\hspace{-0.3mm}+\hspace{-0.3mm} F_2(\phi,X) R 
\hspace{-0.3mm}+\hspace{-0.3mm} F_3(\phi,X) G_{\mu\nu}\phi^{\mu\nu}
\hspace{-0.3mm}+\hspace{-0.3mm} \sum_{I=1}^{5} A_I(\phi,X)L_I^{(2)}
\hspace{-0.3mm}+\hspace{-0.3mm} \sum_{J=1}^{10} B_J(\phi,X)L_J^{(3)}
\bigg],~~~~~
\end{eqnarray}
where 
$g_{\mu\nu}$ is the metric;
$g := {\rm det}(g_{\mu\nu})$;
$R$ is the Ricci scalar associated with $g_{\mu\nu}$;
$\phi$ is the scalar field;
$\phi_\mu:= \nabla_\mu\phi$;
$X:= \phi_\mu\phi^\mu$, the quadratic terms are
\begin{eqnarray}
&& 
L_1^{(2)} =\phi_{\mu\nu}\phi^{\mu\nu}
\quad
L_2^{(2)} =  \left(\Box\phi\right)^2,
\quad
L_3^{(2)} = \Box\phi \phi^\mu \phi_{\mu\nu}\phi^\nu, 
\quad
L_4^{(2)} = \phi^\mu \phi_{\mu\rho}\phi^{\rho\nu}\phi_\nu,
 \quad
L_5^{(2)} = \left(\phi_\mu \phi^{\mu\nu}\phi_{\nu}\right)^2, 
\end{eqnarray}
and the cubic terms are
\begin{eqnarray}
&& 
L_1^{(3)} =\left( \Box\phi\right)^3 
\quad
L_2^{(3)} = \Box\phi \phi_{\mu\nu}\phi^{\mu\nu}, 
\quad
L_3^{(3)} = \phi_{\mu\nu} \phi^{\nu\rho} \phi^\mu{}_\rho , 
\quad
L_4^{(3)} = (\Box\phi)^2 \phi_\mu \phi^{\mu\nu}\phi_\nu,
 \quad
L_5^{(3)} = \Box\phi \phi_\mu \phi^{\mu\nu}\phi_{\nu\rho}\phi^\rho, 
\nonumber\\
&&
L_6^{(3)} = \phi_{\mu\nu}\phi^{\mu\nu}\phi_\rho \phi^{\rho\sigma}\phi_\sigma 
\quad
L_7^{(3)} = \phi_\mu \phi^{\mu\nu}\phi_{\nu\rho}\phi^{\rho\sigma}\phi_\sigma, 
\nonumber\\
&&
L_8^{(3)} = \phi_{\mu} \phi^{\mu\nu} \phi_{\nu\rho}\phi^\rho \phi_\sigma \phi^{\sigma\lambda}\phi_\lambda, 
\quad
L_9^{(3)} = \Box\phi  (\phi_\rho \phi^{\rho\sigma}\phi_\sigma)^2,
 \quad
L_{10}^{(3)} = (\phi_\rho \phi^{\rho\sigma}\phi_\sigma)^3,
\end{eqnarray}
with $\phi_{\mu\nu} := \nabla_\nu\nabla_\mu\phi$;
$\Box\phi:= g^{\mu\nu} \phi_{\mu\nu}$;
$F_i(\phi,X)$ ($i=0,1,2,3$), 
$A_I(\phi,X)$ ($I=1,\cdots, 5$),
and 
$B_J(\phi,X)$ ($J=1,\cdots, 10$)
are functions of $\phi$ and $X$.

In order to eliminate the Ostrogradsky ghosts, 
degeneracy conditions are imposed, which depend on the class of DHOST
~\cite{Langlois:2015cwa,Achour:2016rkg,BenAchour:2016fzp,Crisostomi:2017lbg,Langlois:2017dyl}.
Following Ref.~\cite{BenAchour:2016fzp},
we call
degeneracy conditions for the pure quadratic DHOST theories
($F_3=B_1=B_2=\cdots=B_{10}=0$) 
``Class-${}^2$M'' and ``Class-${}^2$N''
for the cases of $F_2= F_2(\phi)$ and $F_2= F_2(\phi,X)$,
respectively\footnote{Here we take ``minimally coupled'' to mean $F_2(X,\phi)= F_2(\phi)$, because this coupling is free from Ostrogradsky ghosts and does not contribute to the degeneracy conditions, while in Ref. \cite{BenAchour:2016fzp} the authors define ``minimally coupled'' as $F_2=0$.}.
Similarly,
we call 
degeneracy conditions for the pure cubic DHOST theories
($F_2=F_2(\phi)$ and $A_1=A_2=\cdots=A_5=0$) 
``Class-${}^3$M'' and ``Class-${}^3$N''
for the cases of $F_3= 0$ and $F_3\neq 0$,
respectively.
Degeneracy conditions for the pure quadratic, pure cubic, and full DHOST theories
are reviewed in Appendices \ref{app_a}-\ref{app_c}.
We note that the terms $F_0$ and $F_1$ are not relevant for the degeneracy conditions.

We also note that
the pure quartic GLPV ($G_4=G_4(\phi,X)$ and $f_4=f_4(\phi,X)$) and the pure quartic beyond-Horndeski ($G_4=G_4(\phi)$ and $f_4=f_4(\phi,X)$) theories
belong to Class-${}^2$N-I and Class-${}^2$M-I,
and   
the pure quintic GLPV ($G_5=G_5(\phi,X)$ and $f_5=f_5(\phi,X)$) and the pure quintic beyond-Horndeski ($G_5=0$ and $f_5=f_5(\phi,X)$) theories
belong to Class-${}^3$N-I and Class-${}^3$M-I,
respectively,
where we will use Eqs. (2.1)-(2.3) and Eqs. (2.6)-(2.7) in Ref. \cite{Langlois:2018dxi}
for the conventions of the Horndeski and beyond-Horndeski theories
\footnote{
Here,
``pure beyond-Horndeski''
means 
that 
within the quartic or quintic order GLPV theory, only the function $f_4$ or $f_5$
is a nontrivial function of $X$,
while $G_4=G_4(\phi)$ or $G_5=0$ respectively.
}.
In full DHOST theories in which all the terms in Eq. \eqref{qdaction} are nontrivial,
not all the combinations of the pure quadratic and cubic DHOST theories are degenerate, and even if they can be degenerate
some additional conditions may have to  be imposed \cite{BenAchour:2016fzp}.

In the rest of the paper, we study the shift-symmetric subclass of DHOST, where
\be
\label{shift}
 F_i = F_i(X)\quad (i=0,1,2,3), 
\qquad
A_I = A_I(X) \quad (I=1,\cdots, 5),
\qquad 
B_J= B_J(X) \quad
(J=1,\cdots,10).
 \ee
The shift symmetry simplifies the equations of motion dramatically.
With the shift symmetry,
the value of the scalar field itself has no physical meaning, and 
one has to integrate the scalar field 
equations of motion only once.
We also impose the ansatz that the canonical kinetic term is constant
\be \label{eq:constX} X=X_0 =\text{const.} \ee


\subsection{Static and spherically symmetric solutions}

As in Refs. \cite{Babichev:2013cya,Kobayashi:2014wsa,Motohashi:2019sen,Takahashi:2019oxz},
we focus on a static spherically symmetric spacetime 
\be \label{sss} ds^2 = - A(r)\hspace{0.2mm}dt^2 + \f{dr^2}{B(r)} + 2C(r)dt\hspace{0.4mm}dr+ D(r)r^2(d\theta^2+\cos^2\theta \hspace{0.4mm} d\varphi^2) , \ee
and a linearly time-dependent scalar field
\be \label{phi-qt} \phi(t,r)=q\hspace{0.3mm}t+\psi(r).
\ee
 Our choice of linear time-dependence is motivated by the fact that this gives equations of motion which are time-independent (whereas time-dependent equations of motion would be at odds with the static BHs we are investigating).
On the other hand, choosing time-independence (i.e. setting $q=0$) would mean that only a trivial solution ($\phi=$\hspace{0.5mm}constant) is allowed because of the no-hair theorem
\cite{Hui:2012qt,Babichev:2017guv}.

According to the gauge fixing theorem in Ref. \cite{Motohashi:2016prk},
we can set $D(r)=1$ at the level of the action 
without losing any independent equation
of motion, 
while  
choosing $C(r)=0$ at the level of the action
loses an independent equation for $C$.
So that we maintain an independent 
equation, the action must be varied
before imposing $C(r)=0$.
\footnote{This is different to the static scalar field in a static spherically symmetric spacetime, where it is possible to choose both $C(r)=0$ and $D(r)=1$ at the level of the action~\cite{Motohashi:2016prk}.}
We derive the equations of motion by varying the action, and afterwards impose the gauge
conditions $C(r)=0$ and $D(r)=1$.
Once we have fixed the gauge, the kinetic term reads
\be \label{kin}
X(r) = -\f{q^2}{A} + B \psi'^2,
\ee
where a prime denotes a derivative with respect to $r$. 
Using Eq. \eqref{kin}, one can solve Eq. \eqref{eq:constX} for $\psi'$ as
\be \label{psipsol} \psi'=\pm \sqrt{\f{q^2+X_0A}{AB}}, \ee
and with this relation
higher derivatives of $\psi$ can be erased in terms of $q$, $X_0$, $A(r)$, $B(r)$, and derivatives of $A(r)$, $B(r)$.
By taking the positive branch of 
Eq. \eqref{psipsol}
and using ingoing Eddington$-$Finkelstein coordinates $(v, r)$ defined by $dv = dt + dr/\sqrt{AB}$, we obtain $\phi\simeq q\hspace{0.3mm}v$ in the vicinity of
the future event horizon.
Thus, the scalar field is regular at the future event horizon \cite{Babichev:2013cya,Kobayashi:2014eva}.
If we take the negative branch of Eq. \eqref{psipsol},
we obtain the scalar field solution regular at the past event horizon.
From now on, we focus on the BH solutions and 
take the positive branch of Eq. \eqref{psipsol}.
We note that when choosing shift symmetry using Eq. \eqref{shift} and a constant $X$ using Eq. \eqref{eq:constX},
the couplings and their derivatives in the equations of motion evaluated at $X=X_0$ reduce to numbers rather than functions, which makes solving the equations of motion significantly easier.

\section{
The pure cubic DHOST theories} 
\label{sec3}

In this section, we derive the conditions that guarantee the existence
of Schwarzschild or Schwarzschild-(anti-)de Sitter solutions
in the pure cubic theories,
defined by 
\begin{eqnarray}
\label{pure_cubic}
F_2(X)=\frac{M_p^2}{2},
\qquad 
A_I(X)=0
 \quad  (I=1,\cdots, 5).
\end{eqnarray} 
where $M_p$ is the reduced Planck mass,
such that the $F_2$ term reduces to the Einstein-Hilbert term.
We also provide the general form of the coupling functions 
which satisfy the conditions.

\subsection{Schwarzschild solution}
\label{sec3a}

First, let us focus on the Schwarzschild solution
\begin{eqnarray}
\label{schwarzschild}
A=B= 1-\f{r_g}{r},
\end{eqnarray}
where $r_g$ is a constant which corresponds to the location of the event horizon.

We will clarify the conditions
that the metric Eq. \eqref{schwarzschild} and the scalar field Eq. \eqref{phi-qt} with Eq. \eqref{psipsol} 
are the solutions of the scalar-tensor theories Eq. \eqref{qdaction} with
Eq. \eqref{pure_cubic}.
We substitute Eqs. \eqref{schwarzschild} and \eqref{phi-qt} with Eq. \eqref{psipsol}
into the equations of motion to produce  
a Taylor series in $r-r_g$.
We impose that the coefficient which depends on the coupling functions and 
their derivatives evaluated at $X=X_0$
vanishes at each order of $r-r_g$,
and repeat the same procedure until the sufficient number of the conditions are obtained
\cite{Babichev:2016kdt,BenAchour:2018dap,Motohashi:2019sen}. 
As one goes to higher order in $r-r_g$,
the series expansion covers 
a larger domain of $r>r_g$.
After the consistent conditions are obtained,
all the higher order terms of $r-r_g$ trivially vanish
and the entire domain $0<r<\infty$ can be covered.
We note that
our conditions do not depend on the reference position of the Taylor expansion of the equations of motion, and 
the same conditions are obtained
if we expand the same equations with respect to
the other radius, for instance, $r=\infty$.
As in the case of the pure quadratic DHOST theories,
the conditions that the solutions exist are classified into the following two cases:
one is satisfied for $Q:=X_0+q^2=0$,
and 
the other is for $Q\neq 0$.

For $Q:=X_0+q^2=0$ (Case 1-cubic),
the conditions for the existence of the stealth Schwarzschild solutions
are
\begin{eqnarray}
\label{case-1c}
&&
F_0=F_{0X}=F_{1X}=0,
\nonumber\\
&&
B_{3X}
=\frac{3}{10X_0}
\left[
-2B_2 +B_3 
-3 X_0\left(B_4+B_6+2(B_{1X}+B_{2X}) \right)
\right],
\qquad
B_3= -18B_1,
\qquad
B_2=9B_1,
\end{eqnarray}
which are evaluated at $X=X_0=-q^2$. 

For $Q\neq 0$ (Case 2-cubic):
\begin{eqnarray}
\label{case-2c}
&&F_0=F_{0X}=F_{1X}=0,
\nonumber\\
&&
B_1=B_2=B_3=B_{3X}= B_6=0,
\qquad
2B_{1X}+B_{2X}=0,
\qquad
B_4+B_{2X}=0.
\end{eqnarray}
which are evaluated at $X=X_0$.
We note that functions which do not appear in these conditions
are not relevant for the existence of the solutions.
As we will see in Sec. \ref{sec3b},
the constant $F_{0} (X_0)$ serves as the effective cosmological constant,
and should vanish for an asymptotically flat solution.

The general form of the couplings satisfying 
Eqs. \eqref{case-1c} and \eqref{case-2c}
is given by the Taylor series:
\begin{eqnarray}
\label{taylor-c}
F_i(X)=\sum_{j=0}^\infty   f_{ij} (X-X_0)^j,
\qquad
B_J(X)=\sum_{j=0}^\infty  b_{Jj} (X-X_0)^j,
\end{eqnarray}
where $i=0,1,3$
($F_2=M_p^2/2$)
and 
$J=1,2,3,\cdots, 10$.
The relations between the coefficients are given by Eqs.
\eqref{coe-1c} and \eqref{coe-2c}.

\subsection{Schwarzschild-(anti-)de Sitter solution}
\label{sec3b}

Second, let us focus on the Schwarzschild- (anti-) de Sitter solution
\begin{eqnarray}
\label{sch_ds}
A=B= 1-\f{r_g}{r} -\frac{\Lambda}{3}r^2,
\end{eqnarray}
where $r_g$ is a constant which corresponds to the mass of the Schwarzschild- (anti-) de Sitter spacetime
and $\Lambda$ represents the effective
cosmological constant.
We substitute Eqs. \eqref{sch_ds} and \eqref{phi-qt} with Eq. \eqref{psipsol}
into the equations of motion to produce  
an inverse power series of $r$ with respect to $r=\infty$.
We impose that the coefficient which depends on the coupling functions and 
their derivatives evaluated at $X=X_0$
vanishes at each order of $1/r$,
and repeat the same procedure until the sufficient number of the conditions are obtained. 
The conditions can be classified into two cases:
One is satisfied for $Q:=X_0+q^2=0$,
and 
the other is for $Q\neq 0$.

For $Q=0$ (Case 1-$\Lambda$-cubic):
\begin{eqnarray}
\label{case-1lc}
&&
F_0=-\Lambda M_p^2,
\qquad
F_{0X}=0,
\quad
F_{1X}
-\Lambda
\left(
-X_0 B_4+4X_0 B_{1X}+2X_0 B_{2X}+F_{3X}
\right)
=0,
\nonumber\\
&&
B_1
=B_2
=B_3
=B_{3X}
=0,
\qquad
B_4+B_6
+2(B_{1X}+B_{2X})
=0,
\end{eqnarray}
where 
the functions are evaluated at $X=X_0=-q^2$.

For $Q\neq 0$ (Case 2-$\Lambda$-cubic):
\begin{eqnarray}
\label{case-2lc}
&&
F_0=-\Lambda M_p^2,
\qquad
F_{0X}=0,
\qquad
F_{1X}
-\Lambda (-2X_0 B_{1X}+F_{3X})=0,
\nonumber\\
&&
B_1
=B_2
=B_3
=B_{3X}
=B_6
=0,
\qquad
B_{2X}+2B_{1X}=0,
\qquad
B_4-2B_{1X}=0,
\end{eqnarray}
where 
the functions are evaluated at $X=X_0$.
We note that the functions which do not appear in these conditions
are not relevant for the existence of the solutions.
The general form of the couplings satisfying 
Eqs. \eqref{case-1lc} and \eqref{case-2lc}
is given by the Taylor series Eq. \eqref{taylor-c},
with the relations between coefficients given by
\eqref{coe-1lc} and \eqref{coe-2lc}.

Finally,
we should note that 
for both the cases of the Schwarzschild and Schwarzschild- (anti-) de Sitter cases,
there would also be solutions with solid deficit angles
with $B=\alpha \left(1-r_g/r -\Lambda r^2/3\right)$ ($\alpha\neq 1$) 
as in the pure quadratic DHOST theories \cite{Motohashi:2019sen,Takahashi:2019oxz}.
These solutions are very similar to the global monopole solutions obtained in \cite{Barriola:1989hx}.
Since in contrast to the case of the global monopole,
solutions with deficit angles are not the unique solutions in our theory
and not physically important in comparison with the exact
Schwarzschild or Schwarzschild-(anti-) de Sitter solutions.
Thus in this paper
we will not consider these solutions.

\subsection{Compatibility with the degeneracy conditions}

Now, we impose degeneracy conditions of pure cubic DHOST theories shown in Appendix \ref{app_b},
and check their compatibility with the conditions \eqref{case-1c}, \eqref{case-2c}, \eqref{case-1lc}, and \eqref{case-2lc}.
In
Table \ref{table1}, we show which class of pure cubic DHOST theories defined in Appendix. \ref{app_b}
are compatible with the different BH solutions, recalling 
that Case 1(-$\Lambda$)-cubic and 2(-$\Lambda$)-cubic refer to $X_0=-q^2$ and $X_0\neq -q^2$ respectively:
\begin{table}[ht]
\begin{tabular}{ c| c c c c }
& Case 1-cubic & Case 2-cubic & Case 1-$\Lambda$-cubic & Case 2-$\Lambda$-cubic \\ 
 \hline
${}^{3}$M-I & \checkmark & $\times$ & $\times$ & $\times$ \\  
${}^{3}$M-II &  $\times$ & $\times$ & $\times$ & $\times$\\ 
${}^{3}$M-III & $\times$ & $\times$ & $\times$ & $\times$ \\  
${}^{3}$M-IV & \checkmark\footnotemark[1] & \checkmark & \checkmark & \checkmark   \\ 
${}^{3}$M-V & $\times$ & $\times$ & $\times$ & $\times$ \\  
${}^{3}$M-VI & \checkmark\footnotemark[1] & \checkmark  & \checkmark & \checkmark \\  
${}^{3}$M-VII & \checkmark\footnotemark[1] & \checkmark & \checkmark & \checkmark \\
\hline
${}^{3}$N-I & $\times$ & $\times$  & $\times$ & $\times$  \\  
${}^{3}$N-II & \checkmark \footnotemark[1]& \checkmark & \checkmark & \checkmark   
\end{tabular}
\caption{Compatibility of degeneracy conditions
in the cubic DHOST theories with  the conditions in Case 1(-$\Lambda$)-cubic and Case 2(-$\Lambda$)-cubic.
}\label{table1}
\footnotetext[1]{The conditions of Case 1-cubic
are compatible with Classes-${}^3$M-IV, ${}^3$M-VI, ${}^3$M-VII, and ${}^{3}$N-II
only for $B_1=B_2=B_3=0$ because of the different ratios of $B_2$ and $B_3$ to $B_1$. }
\end{table}

Thus
the conditions of the stealth Schwarzschild solutions 
are compatible with 
Classes-${}^3$M-I, ${}^3$M-IV, ${}^3$M-VI, ${}^3$M-VII, and $^3$N-II; 
the conditions of the Schwarzschild- (anti-) de Sitter solutions 
are compatible
with 
Classes-${}^3$M-IV, ${}^3$M-VI, ${}^3$M-VII, and $^3$N-II,
respectively.

\section{
The full DHOST theories}
\label{sec4}

In this section, we derive the conditions that guarantee the existence
of Schwarzschild or Schwarzschild- (anti-)de Sitter solutions
in the full DHOST theories, namely,
in the presence of both quadratic and cubic DHOST couplings
$F_i(X)\neq 0$, 
$A_I(X)\neq 0$,
and 
$B_J(X)\neq  0$,
where
$i=0,1,2,3;
I=1,\cdots, 5;\, J=1,\cdots,10$,
and provide the general form of the coupling functions which satisfy these conditions. We find that there is no mixing between quadratic and cubic terms, i.e. there are no conditions containing both $A_I$s and $B_J$s.

\subsection{Schwarzschild solution}
\label{sec4a}

First, let us focus on the Schwarzschild solution given by the metric Eq. \eqref{schwarzschild}.
The procedure for obtaining the solution 
is the same as that in Sec. \ref{sec3a}, i.e. we substitute the metric Eq. \eqref{schwarzschild} and the scalar field Eq. \eqref{phi-qt} with Eq. \eqref{psipsol} into the equations of motion and
find the conditions such that the coefficients of the Taylor series of the equations of motion vanish.
Here, we only focus on the result.

The conditions that the solutions exist are as follows:

For $Q=X_0+q^2=0$ (Case 1-full):
\begin{eqnarray}
\label{case-1f}
&&
F_0=F_{0X}=F_{1X}=0,
\nonumber\\
&&
A_1+A_2=0,
\qquad
A_{1X}+A_{2X}=0,
\nonumber\\
&&
B_{3X}
=\frac{3}{10X_0}
\left[
-2B_2 +B_3 
-3 X_0\left(B_4+B_6+2(B_{1X}+B_{2X}) \right)
\right],
\qquad
B_3= -18B_1,
\qquad
B_2=9B_1,
\end{eqnarray}
which are evaluated at $X=X_0=-q^2$. 
We note that 
the limit to pure quadratic DHOST theories
($F_3=0$ and $B_1=B_2=\cdots =B_{10}=0$)
agrees with Case 1 obtained in Ref. \cite{Motohashi:2019sen}.

For $Q\neq 0$ (Case 2-full),
the conditions for the existence of the stealth Schwarzschild solutions
are given by 
\begin{eqnarray}
\label{case-2f}
&&F_0=F_{0X}=F_{1X}=0,
\nonumber\\
&&
A_1=A_2=0,
\qquad
A_{1X}+A_{2X}=0,
\qquad
A_3+2A_{1X}=0,
\nonumber\\
&&
B_1=B_2=B_3=B_{3X}= B_6=0,
\qquad
2B_{1X}+B_{2X}=0,
\qquad
B_4+B_{2X}=0.
\end{eqnarray}
which are evaluated at $X=X_0$.
We note that 
the limit to pure quadratic DHOST theories
($F_3=0$ and $B_1=B_2=\cdots =B_{10}=0$)
agrees with Case 2 obtained in Ref. \cite{Motohashi:2019sen}. 
Thus, for the existence of the Schwarzschild solution,
the conditions for the pure quadratic and pure cubic DHOST theories
can be independently satisfied,
and no extra conditions with the mixture of $A_I$s and $B_J$s, 
and their derivatives, are imposed.

The general form of the couplings satisfying 
Eqs. \eqref{case-1f} and \eqref{case-2f}
is given by the Taylor series
\begin{eqnarray}
\label{taylor-f}
F_i(X)=\sum_{j=0}^\infty   f_{ij} (X-X_0)^j,
\qquad 
A_I(X)=\sum_{j=0}^\infty   a_{Ij} (X-X_0)^j,
\qquad 
B_J(X)=\sum_{j=0}^\infty  b_{Jj} (X-X_0)^j,
\end{eqnarray}
where $i=0,1,2,3$,
$I=1,2,\cdots, 5$,
and
$J=1,2,3,\cdots, 10$,
with the relations between coefficients given in \eqref{coe-1f} and \eqref{coe-2f},
respectively.

\subsection{Schwarzschild- (anti-) de Sitter solution}
\label{sec4b}

Second, let us focus on the Schwarzschild-(anti-) de Sitter solution given by the metric Eq. \eqref{sch_ds}, i.e. we substitute the metric and the scalar field Eq. \eqref{phi-qt} with Eq. \eqref{psipsol} into the equations of motion and
find the conditions such that the coefficients of the Taylor series of the equations of motion vanish.
The procedure for obtaining the solution 
is the same as that in Sec. \ref{sec3b}.
Here, we only focus on the result.

The conditions that we get the solutions are as follows:

For $Q:=X_0+q^2=0$ (Case 1-$\Lambda$-full):
\begin{eqnarray}
\label{case-1lf}
&&
F_0+2\Lambda F_2=2\Lambda X_0 A_1,
\quad
F_{0X}+4\Lambda F_{2X}=
\frac{\Lambda}{2}
(2A_1-3X_0 A_3-4X_0 A_{1X}),
\nonumber\\
&&
F_{1X}
-\Lambda
\left(
-X_0 B_4+4X_0 B_{1X}+2X_0 B_{2X}+F_{3X}
\right)
=0,
\nonumber\\
&&
A_1+A_2=0,
\qquad
A_{1X}+A_{2X}=0,
\nonumber\\
&&
B_1
=B_2
=B_3
=B_{3X}
=0,
\qquad
B_4+B_6
+2(B_{1X}+B_{2X})
=0,
\end{eqnarray}
where 
the functions are evaluated at $X=X_0=-q^2$.
We note that 
the limit to pure quadratic DHOST theories
($F_3=0$ and $B_1=B_2=\cdots =B_{10}=0$)
agrees with Case 1-$\Lambda$ obtained in Ref. \cite{Motohashi:2019sen}.

For $Q\neq 0$ (Case 2-$\Lambda$-full):
\begin{eqnarray}
\label{case-2lf}
&&
F_{0}+2\Lambda F_{2}=0,
\qquad
F_{0X}
+\Lambda \left(4F_{2X}-X_0 A_{1X}\right)=0,
\qquad
F_{1X}
-\Lambda (-2X_0 B_{1X}+F_{3X})=0,
\nonumber\\
&&
A_{1}=A_2=0,
\qquad
A_{1X}+A_{2X}=0,
\qquad
A_{3}+2A_{1X}=0,
\nonumber\\
&&
B_1
=B_2
=B_3
=B_{3X}
=B_6
=0,
\qquad
B_{2X}+2B_{1X}=0,
\qquad
B_4-2B_{1X}=0,
\end{eqnarray}
where 
the functions are evaluated at $X=X_0$.
We note that 
the limit to pure quadratic DHOST theories
($F_3=0$ and $B_1=B_2=\cdots =B_{10}=0$)
agrees with Case 2-$\Lambda$ obtained in Ref. \cite{Motohashi:2019sen}.

Thus, for the existence of the Schwarzschild- (anti-) de Sitter solution,
the conditions for the pure quadratic and pure cubic DHOST theories
have to be independently satisfied,
and no extra conditions with the mixture of $A_I$s and $B_J$s, 
and their derivatives, are imposed.
The general form of the couplings satisfying 
Eqs. \eqref{case-1lf} and \eqref{case-2lf}
is given by the Taylor series Eq. \eqref{taylor-f},
with the relations between coefficients given by
\eqref{coe-1lf} and \eqref{coe-2lf}.

As we noted in the case of the pure cubic DHOST theories,  
for both the cases of the Schwarzschild and Schwarzschild-(anti-) de Sitter cases,
there would also be solutions with solid deficit angles.
Again, 
these solutions are not physically important in comparison with the exact
Schwarzschild or Schwarzschild-(anti-) de Sitter solutions,
and therefore we focus on these solutions.

\subsection{Compatibility with the degeneracy conditions}

We then impose degeneracy conditions of the full DHOST theories shown in Appendix \ref{app_c},
and check the compatibility of the conditions \eqref{case-1f}, \eqref{case-2f}, \eqref{case-1lf}, and \eqref{case-2lf} with them.
Combining all conditions, we find that only a few combinations
of quadratic and cubic theories allow BH solutions.
Our results are summarized in
Table \ref{table2}, which shows the 
combinations of theories which allow both Schwarzschild and Schwarzschild-(anti-) de Sitter solutions.
\footnote{The combination of Class-${}^3$M-I 
with
Class-${}^2$M-I
and 
that of 
Class-${}^3$M-I 
with
Class-${}^2$N-I
permit
Schwarzschild solutions as long as $X_0=-q^2$, but not de Sitter-Schwarzschild solutions. In other words they only permit Case-I-full solutions.} 
We note that 
the additional condition $A_1=A_2=0$ is required 
for both of these combinations when looking for Case 2-full and Case 2-$\Lambda$-full theories.
\begin{table}[ht]
\begin{tabular}{ c| c c c c c c c | c c}
 Class& ${}^3$M-I &${}^3$M-II & ${}^3$M-III & ${}^3$M-IV &
${}^3$M-V & ${}^3$M-VI & ${}^3$M-VII & ${}^3$N-I & ${}^3$N-II\\ 
 \hline
${}^2$M-I & Case-1-full only & $\times$ & $\times$ & $-$ & $\times$ &  $-$ & $-$ & $-$ & $-$ \\  
${}^2$M-II\footnotemark[1]& $-$  & $-$  & $\times$ &  \checkmark & $-$ & $-$ & \checkmark &  $-$& $-$ \\ 
${}^2$M-III\footnotemark[1]&$-$ & $-$ & $-$ & $-$ & $\times$ & \checkmark & \checkmark & $-$ & $-$\\  
 \hline
${}^2$N-I & Case 1-full only & $\times$ & $\times$ &  $-$& $\times$ & $-$ & $-$& $\times$ & $-$ \\  
${}^2$N-II & $-$ & $-$ & $-$ & $-$& $-$ & $-$ & \checkmark& $-$ &  \checkmark\\ 
${}^2$N-III & $\times$ & $\times$ & $\times$ & $-$ & $\times$ & $-$ & $-$& $-$ & $-$\\  
${}^2$N-IV & $\times$  & $\times$  & $\times$& $-$ & $\times$ & $-$& $-$ & $-$ & $-$    
\end{tabular}
\caption{Compatibility of degeneracy conditions
in full DHOST theories with the conditions in Case 1(-$\Lambda$)-full and Case 2(-$\Lambda$)-full. The combinations denoted by ``$-$''
were already excluded in terms of the incompatibility between degeneracy conditions,
while
the combinations denoted by ``$\times$''
are ruled out in terms of the incompatibility of degeneracy conditions with
the conditions for the existence of Schwarzschild and Schwarzschild- (anti-) de Sitter BH solutions.}
\footnotetext[1]{Class ${}^2$M-II has the additional condition $A_1=A_2=A_3=0$ and ${}^2$M-III
has the condition $A_1=A_2=0$ for all solutions.}
\label{table2}
\end{table}

Since
the combination of
the pure quartic order Horndeski theory ($G_4\neq 0$ while $f_4= 0$) among Class-${}^2$N DHOST and the pure quintic order beyond-Horndeski theory ($f_5\neq 0$ while $G_5=0$)
among Class-${}^3$M DHOST cannot be degenerate, Case 1-full solutions do not exist in this limit (See Sec. \ref{sec5}).

\section{GLPV and Horndeski theories}
\label{sec5}

Particular choices of the quadratic and cubic DHOST theories
correspond to the Horndeski and GLPV theories.

The Horndeski theories corresponds to the choice
\begin{eqnarray}
&&
F_0=G_2,
\qquad
F_1=G_3,
\qquad 
F_2=G_4,
\qquad
F_3=G_5,
\nonumber\\
&&
A_1=-A_2 =2G_{4X},
\qquad
A_3=A_4=A_5=0,
\nonumber\\
&&
3B_1
=-B_2
=\frac{3}{2}B_3
=G_{5X},
\qquad
B_J=0
\,\,\,
(J=4, \cdots,10).
\end{eqnarray}
Similarly, 
the GLPV theories correspond to the choice
\begin{eqnarray}
\label{eq:GLPVchoice}
&& F_0=G_2,
\qquad
F_1=G_3,
\qquad
F_2=G_4,
\qquad
F_3=G_5,
\nonumber\\
&&
A_1=-A_2
=2G_{4X}+Xf_4,
\qquad
A_3=-A_4
=2f_4,
\qquad
A_5=0,
\nonumber\\
&&
3B_1
=-B_2
= \frac{3}{2} B_3
=G_{5X}
+3X f_{5},
\qquad
-2B_4
=B_5
=2B_6
=-B_7
=6f_5,
\qquad 
B_8
=B_9
=B_{10}
=0.
\end{eqnarray}
The case of \eqref{eq:GLPVchoice} with $f_4=0$ and $f_5= 0$
corresponds to Horndeski theories. 
The pure quartic order GLPV theory ($G_4\neq 0$ and $f_4\neq 0$ while $G_5=f_5=0$)
belongs to Class-${}^2$N-I DHOST,
and the pure quintic order GLPV theory ($G_5\neq 0$ and $f_5\neq 0$ while $G_4=M_p^2/2$ and $f_4=0$)
belongs to Class- ${}^3$N-I DHOST
\cite{BenAchour:2016fzp}.
We note that 
the combination of the quartic order Horndeski term ($G_4$) and the pure quintic order beyond-Horndeski term ($f_5$),
and 
that of the pure quintic order beyond-Horndeski term ($f_5$) and the quartic order Horndeski term ($G_4$),
are both nondegenerate \cite{Langlois:2015cwa,Crisostomi:2016tcp}. 
On the other hand,
the combination of the quartic order Horndeski ($G_4$) and pure beyond-Horndeski ($f_4$) terms,
and 
that of the quintic order Horndeski term ($G_5$) and pure beyond-Horndeski ($f_5$) terms
are degenerate \cite{Langlois:2015cwa,Crisostomi:2016tcp}.

\subsection{The pure cubic DHOST theories}

\subsubsection{GLPV theories}

The GLPV limit of the pure cubic DHOST theories
corresponds to the choice of $G_4(X)=M_p^2/2$ and $f_4(X)=0$.
In the GLPV limit, 
the conditions in Case 1-cubic reduce to 
\begin{eqnarray}
&&
G_2
=G_{2X}
=G_{3X}
=
G_{5X}+3X_0 f_5
=
G_{5XX} 
+3f_5+3X_0 f_{5X}
=0,
\end{eqnarray}
evaluated at $X=X_0=-q^2$.
The conditions in Case 2-cubic reduce to 
\begin{eqnarray}
&&
G_2
=G_{2X}
=G_{3X} 
=
G_{5X}
=f_5
= G_{5XX}+3X_0 f_{5X}
=0,
\end{eqnarray}
evaluated at $X=X_0$.

The conditions in Case 1-$\Lambda$-cubic reduce to 
\begin{eqnarray}
&&
G_2+\Lambda M_p^2
=G_{2X} 
=G_{3X}
=
G_{5X}+3X_0 f_5
=
G_{5XX}+3 (f_5+X_0 f_{5X})
=0,
\end{eqnarray}
evaluated at $X=X_0=-q^2$.
The conditions in Case 2-$\Lambda$-cubic reduce to 
\begin{eqnarray}
&&
  G_2+\Lambda M_p^2
= G_{2X}
=G_{3X}
=G_{5X}
=f_5
= G_{5XX}+3X_0 f_{5X}
=0,
\end{eqnarray}
evaluated at $X=X_0$.

\subsubsection{Horndeski limit}
In the case of the Horndeski theories,
$f_5(X)=0$, 
the conditions for Case 1-cubic and Case 2-cubic
further reduce to 
\begin{eqnarray}
&&
G_2
=G_{2X}
=G_{3X}
=G_{5X}
=G_{5XX} 
=0,
\end{eqnarray}
evaluated at $X=X_0=-q^2$,
and 
$X=X_0\neq-q^2$,
respectively.
In the Horndeski limit, the conditions for Case 1$-\Lambda$-cubic and Case 2$-\Lambda$-cubic
further reduce to 
\begin{eqnarray}
&&
G_2
+\Lambda M_p^2
=
G_{2X}
=
G_{3X}
=
G_{5X}
=
G_{5XX}
=0,
\end{eqnarray}
evaluated at $X=X_0=-q^2$
and 
$X=X_0\neq -q^2$, respectively.

\subsection{The full DHOST theories}

\subsubsection{GLPV theories}

In the GLPV limit, 
the conditions in Case 1-full reduce to 
\begin{eqnarray}
&&
G_2
=G_{2X}
=G_{3X}
=
G_{5X}+3X_0 f_5
=
G_{5XX} 
+3f_5+3X_0 f_{5X}
=0,
\end{eqnarray}
evaluated at $X=X_0=-q^2$.
The conditions in Case 2-full reduce to 
\begin{eqnarray}
&&
G_2
=G_{2X}
=G_{3X} 
=
2G_{4X}
+X_0 f_4
=
2G_{4XX}
+X_0 f_{4X}
+2f_4
=
G_{5X}
=f_5
= G_{5XX}+3X_0 f_{5X}
=0,
\end{eqnarray}
evaluated at $X=X_0$.

The conditions in Case 1-$\Lambda$-full reduce to 
\begin{eqnarray}
&&
G_2+2\Lambda G_4
-2\Lambda X_0
\left(2G_{4X} +X_0 f_{4}\right)
=
G_{2X}
+
2\Lambda
\left(
 2X_0f_4 
+X_0^2f_{4X}
+G_{4X}
+2X_0G_{4XX}
\right)
\nonumber\\
&=&
G_{3X}
=
G_{5X}+3X_0 f_5
=
G_{5XX}+3 (f_5+X_0 f_{5X})
=0,
\end{eqnarray}
evaluated at $X=X_0=-q^2$.
The conditions in Case 2-$\Lambda$-full reduce to 
\begin{eqnarray}
&&
2G_{4X}+X_0 f_4
=2G_{4XX}+X_0 f_{4X}+2f_4
= G_2+2\Lambda G_4
= G_{2X}+\Lambda(4 G_{4X}+X_0 f_4)
\nonumber\\
&=&
G_{3X}
=G_{5X}
=f_5
= G_{5XX}+3X_0 f_{5X}
=0,
\end{eqnarray}
evaluated at $X=X_0$.
These results agree with those in Ref. \cite{Motohashi:2019sen} which takes $G_5(X)=f_5(X)=0$,
and the particular subclass of the quartic order GLPV theories which provides the self-tuning mechanism of the cosmological constant \cite{Babichev:2016kdt,Babichev:2017lmw}.

In the case that both the quartic- and quintic- order beyond-Horndeski interactions ($f_4(X)\neq 0$ and $f_5(X)\neq 0$) are present,
the theory can be degenerate only if $G_4(X)=G_5(X)=0$ \cite{Langlois:2015cwa,Crisostomi:2016tcp}.
Then, 
the conditions in Case 1-full and Case 2-full further reduce to 
\begin{eqnarray}
&&
G_2
=G_{2X}
=G_{3X}
= f_5
= f_{5X}
=0,
\end{eqnarray}
evaluated at $X=X_0=-q^2$,
and 
\begin{eqnarray}
&&
G_2
=G_{2X}
=G_{3X} 
=f_4
=f_{4X}
=f_5
=f_{5X}
=0,
\end{eqnarray}
evaluated at $X=X_0\neq-q^2$, respectively.
The conditions in Case 1-$\Lambda$-full and Case 2-$\Lambda$-full
further reduce to 
\begin{eqnarray}
&&
G_2
-2\Lambda X_0^2 f_{4X}
=
G_{2X}
+
2\Lambda
\left(
 2X_0f_4 
+X_0^2f_{4X}
\right)
=
G_{3X}
=
 f_5
=
f_{5X}
=0,
\end{eqnarray}
evaluated at $X=X_0=-q^2$,
and 
\begin{eqnarray}
&&
 f_4
=
 f_{4X}
= G_2
= G_{2X}
=
G_{3X}
=f_5
= f_{5X}
=0,
\end{eqnarray}
evaluated at $X=X_0\neq -q^2$,
respectively.
All these conditions are not compatible with the assumption
that $f_4(X)\neq 0$ and $f_5(X)\neq 0$. Thus, 
the existence of the Schwarzschild and Schwarzschild-(anti-) de Sitter solutions
cannot be compatible with degeneracy of GLPV theories
with both the quartic and quintic beyond-Horndeski interactions.

\subsubsection{Horndeski limit}
In the case of the Horndeski theories,
$f_4(X)=f_5(X)=0$, 
the conditions for Case 1-full and Case 2-full
further reduce to 
\begin{eqnarray}
&&
G_2
=G_{2X}
=G_{3X}
=G_{5X}
=G_{5XX} 
=0,
\end{eqnarray}
evaluated at $X=X_0=-q^2$,
and 
\begin{eqnarray}
&&
G_2
=G_{2X}
=G_{3X} 
=G_{4X}
=G_{5X}
= G_{5XX}
=0,
\end{eqnarray}
evaluated at $X=X_0$, respectively.
In the Horndeski limit,
the conditions for Case 1$-\Lambda$-full and Case 2$-\Lambda$-full
further reduce to 
\begin{eqnarray}
&&
G_2+2\Lambda G_4
-4\Lambda X_0 G_{4X}
=
G_{2X}
+
2\Lambda
\left(
G_{4X}
+2X_0G_{4XX}
\right)
=
G_{3X}
=
G_{5X}
=
G_{5XX}
=0,
\end{eqnarray}
evaluated at $X=X_0=-q^2$,
and 
\begin{eqnarray}
&&
G_{4X}
=G_{4XX}
= G_2+2\Lambda G_4
= G_{2X}+4\Lambda G_{4X}
=G_{3X}
=G_{5X}
= G_{5XX}
=0,
\end{eqnarray}
evaluated at $X=X_0$, respectively.

These conditions coincide with those obtained in Refs.~\cite{Babichev:2013cya,Kobayashi:2014eva} which takes $G_3(X)=G_5(X)=0$.


\section{Conclusions}
\label{sec6}

We obtained conditions for the existence of the Schwarzschild and Schwarzschild-(anti-) de Sitter black hole solutions in the shift-symmetric DHOST theories by assuming that the scalar field has a nontrivial profile with a linear time dependence and that the ordinary kinetic term of the scalar field is constant. Our analysis extended the work of Ref. \cite{Motohashi:2019sen}, exhausted all the known classes of DHOST theories,
and included the shift-symmetric subclasses of Horndeski and GLPV theories as special cases.

We have also investigated the compatibility between the degeneracy conditions 
and the conditions to allow the exact BH solutions obtained in this paper.
We found
that 
in most classes of DHOST theories with both quadratic and cubic interactions,
these conditions are not compatible,
and cubic DHOST theories are not the scalar-tensor theories giving rise to viable BH solutions with a non-trivial scalar field
which are absent in GR.
Combined with the fact
that none of the cubic DHOST theories satisfy the latest bound on
the propagation speed of gravitational waves \cite{Langlois:2017dyl},
our analysis means that cubic DHOST theories
are disfavoured
in all the regimes of our Universe.
In other words,
the only DHOST theories
which are likely to be viable alternatives to GR 
are quadratic DHOST theories.
Ref. \cite{deRham:2019gha} recently
argued
that even-mode perturbations about
stealth Schwarzschild solutions in pure quadratic DHOST theories suffer strong coupling problems,
indicating that predictions at the level of linear perturbations cannot be trusted.
If this is true, 
the same issue may also be present for the case of the cubic DHOST theories.
All these facts would strongly disfavor DHOST theories in terms of the properties of BHs, in comparison with those in GR.

\acknowledgments{
M.M. was supported by the research grant under ``Norma Transit\'oria do DL 57/2016''.
J.E. acknowledges the support by CENTRA.}


\appendix

\section{The pure quadratic DHOST theories}
\label{app_a}

In the Appendices \ref{app_a}-\ref{app_c},
we summarize the degeneracy conditions 
for the pure quadratic, pure cubic, and full DHOST theories.
In the case of the pure quadratic DHOST theories, $F_3=B_J=0$ (where $J=1,2,\cdots, 10$). 


\subsection{The case $F_2=F_2(\phi)$}

In the case of $F_2= F_2(\phi)$,
degeneracy conditions for minimally coupled quadratic theories\footnote{Here we take ``minimally coupled'' to mean $F_2(X,\phi)= F_2(\phi)$, because this coupling is free from Ostrogradsky ghosts and does not contribute to the degeneracy conditions.} are classified into the following three classes \cite{Crisostomi:2016czh}:

The Class-${}^2$M-I DHOST theory is given by 
\begin{eqnarray}
\label{2m1}
A_4=-\frac{2A_1}{X},
\qquad
A_5
=\frac{4A_1(A_1+2A_2)-4A_1A_3 X+3A_3^2X^2}
        {4(A_1+3A_3)X^2}.
\end{eqnarray}
The pure quartic beyond-Horndeski theories ($f_4\neq 0$ while $G_4=G_4(\phi)$ and $G_5=f_5=0$) satisfy the degeneracy conditions of the Class-${}^2$M-I DHOST
\cite{Langlois:2015cwa,BenAchour:2016fzp}.

The Class-${}^2$M-II DHOST theory is given by 
\begin{eqnarray}
\label{2m2}
A_2=-\frac{A_1}{3},
\qquad
A_3=\frac{2A_1 }{3X}.
\end{eqnarray}

The Class-${}^2$M-III DHOST theory is given by
\begin{eqnarray}
\label{2m3}
A_1
=
0.
\end{eqnarray}

Class-${}^2$M theories are also  called ``Class III'' theories
\cite{Achour:2016rkg,Langlois:2018dxi}.

\subsection{The case $F_2=F_2(\phi,X)$}

In the case of $F_2=F_2(\phi,X)$,
degeneracy conditions are classified into the following four classes \cite{BenAchour:2016fzp}:

The Class-${}^2$N-I DHOST theory with $F_2\neq XA_1$ is given by 
\begin{eqnarray}
\label{2n1}
A_2&=&-A_1\neq -\frac{F_2}{X},
\nonumber\\
A_4&=&\frac{1}{8(F_2-XA_1)^2}
\left\{
4F_2
\left[ 
3(A_1-2F_{2X})^2-2A_3F_2\right]
-A_3X^2(16A_1F_{2X}+A_3F_2) 
\right.
\nonumber\\
&&
\left.
+
4X
\left[
3A_1A_3F_2+16A_1^2F_{2X}-16A_1F_{2X}^2-4A_1^3+2A_3F_2F_{2X}
\right]
\right\}, 
\nonumber\\
A_5&=&\frac{1}{8(F_2-XA_1)^2} (2A_1-XA_3-4F_{2X}) \left(A_1(2A_1+3XA_3-4F_{2X})-4A_3F_2\right).
\end{eqnarray}
The pure quartic order GLPV theories ($G_4\neq 0$ and $f_4\neq 0$ while $G_5=f_5=0$) satisfy the degeneracy conditions of the Class-${}^2$N-I DHOST
\cite{Langlois:2015cwa,BenAchour:2016fzp}.

The Class-${}^2$N-II DHOST theory with $F_2\neq XA_1$ is given by  
is given by 
\begin{eqnarray}
\label{2n2}
A_2=-A_1=-\frac{F_2}{X},
\qquad
A_3=\frac{2(F_2-2XF_{2X}) }{X^2}.
\end{eqnarray}
Class-${}^2$N-I and Class-${}^2$N-II theories which satisfy $A_1+A_2=0$
are also  called ``Class I'' theories
\cite{Achour:2016rkg,Langlois:2018dxi}.

The Class-${}^2$N-III DHOST theory with $A_1+A_2\neq 0$ and $F_2\neq XA_1$ is given by
\begin{eqnarray}
\label{2n3}
A_3
&=&
\frac{4F_{2X} (A_1+3A_2)}{F_2}
-\frac{2(A_1+4A_2-2F_{2X})}{X}
-\frac{4F_2}{X^2},
\nonumber\\
A_4
&=&
\frac{2F_2}{X^2}
+\frac{8F_{2X}^2}{F_{2}}
-\frac{2(A_1+2F_{2X})}{X},
\nonumber\\
A_5
&=&
\frac{2}{F_2^2X^3}
\left[
4F_2^3+F_2^2X (3A_1+8A_2-12F_{2X})
+8F_2 F_{2X}X^2
  (F_{2X}-A_1-3A_2)
+6F_{2X}^2 X^3(A_1+3A_2)
\right].
\end{eqnarray}

The Class-${}^2$N-IV DHOST theory with $A_1+A_2\neq 0$
is given by 
\begin{eqnarray}
\label{2n4}
A_1
&=&
\frac{F_2}{X},
\nonumber\\
A_4
&=&
\frac{8F_{2X}^2}{F_2}
-\frac{4F_{2X}}{X},
\nonumber\\
A_5
&=&
\frac{1}
      {4F_2 X^3(F_2+A_2X)}
\left[
  F_2 A_3^2X^4
-4F_2^3
-8F_2^2 X (A_2-2F_{2X})
\right.
\nonumber\\
&&
\left.
-4F_2X^2 
 \left(4F_{2X} (F_{2X}-2A_2)+A_3F_2\right)
+8F_{2X} X^3 (A_3F_2-4A_2F_{2X})
\right].
\end{eqnarray}

Class-${}^2$N-III and Class-${}^2$N-IV theories which do not satisfy $A_1+A_2=0$
are also called ``Class II'' theories
\cite{Achour:2016rkg,Langlois:2018dxi}.

\section{The pure cubic DHOST theories}
\label{app_b}

In the case of pure cubic DHOST theories, $F_2=A_I=0$ (where $I=1,2,\cdots, 5$).

\subsection{The minimally coupled case $F_3=0$}

In the minimally coupled case $F_3= 0$,
degeneracy conditions are classified into seven classes \cite{BenAchour:2016fzp}:

The Class-${}^3$M-I DHOST theory ($9B_1+2B_2\neq 0$)
 is given by 
 \begin{eqnarray} &&B_{5}=-\frac{2}{X} B_{2}, \quad B_{6}=\frac{9 B_{1} B_{3}+3 B_{4} X\left(B_{2}+B_{3}\right)-2 B_{2}^{2}}{X\left(9 B_{1}+2 B_{2}\right)}, \nonumber\\ 
 && B_{7}=-\frac{3}{X} B_{3}, \quad B_{8}=\frac{9 B_{1} B_{3}-6 B_{4} X\left(B_{2}+B_{3}\right)+6 B_{2} B_{3}+4 B_{2}^{2}}{X^{2}\left(9 B_{1}+2 B_{2}\right)}, \nonumber\\ 
 && B_{9}= \frac{1}{X^{2}\left(9 B_{1}+2 B_{2}\right)^{2}}\left[3 B_{4}^{2} X^{2}\left(9 B_{1}+3 B_{2}+B_{3}\right)-2 B_{4} X\left(9 B_{1}\left(B_{2}-B_{3}\right)+4 B_{2}^{2}\right)
\right.\nonumber\\ 
 &&+24 B_{1} B_{2}^{2}+54 B_{1}^{2} B_{2}+27 B_{1}^{2} B_{3}+4 B_{2}^{3} ] ,
 \nonumber
 \\
 &&B_{10}= \frac{1}{X^{3}\left(9 B_{1}+2 B_{2}\right)^{3}}\left[3 B_{4}^{3} X^{3}\left(9 B_{1}+3 B_{2}+B_{3}\right)-6 B_{2} B_{4}^{2} X^{2}\left(9 B_{1}+3 B_{2}+B_{3}\right)\right.\nonumber\\ 
 &&+2 B_{4} X\left(81 B_{1}^{2}\left(B_{2}+B_{3}\right)+18 B_{1} B_{2}\left(3 B_{2}+2 B_{3}\right)+2 B_{2}^{2}\left(5 B_{2}+3 B_{3}\right)\right) \nonumber\\ 
 &&-2\left(54 B_{1}^{2} B_{2}\left(B_{2}+2 B_{3}\right)+4 B_{1} B_{2}^{2}\left(7 B_{2}+9 B_{3}\right)+81 B_{1}^{3} B_{3}+4 B_{2}^{3}\left(B_{2}+B_{3}\right)\right) ]. \end{eqnarray}
The pure quartic order beyond-Horndeski interactions ($f_4\neq 0$ while $G_4=G_4(\phi)$
and $G_5=f_5=0$) satisfy the degeneracy conditions of Class-${}^3$M-I DHOST
\cite{Langlois:2015cwa,BenAchour:2016fzp}.

The Class-${}^3$M-II DHOST theory ($9B_1-2B_2\neq 0$)
 is given by 
 \begin{eqnarray}
 &&B_{2}=-\frac{9}{2} B_{1}, \quad B_{4}=-\frac{3}{X} B_{1}, \quad B_{5}=\frac{9}{X} B_{1}, \nonumber\\ 
 &&B_{7}=-\frac{3}{X} B_{3}, \qquad B_{8}=\frac{3 B_{3}-2 B_{6} X}{X^{2}}, \nonumber\\ 
 &&B_{9}=\frac{9 B_{1}\left(B_{3}+2 B_{6} X\right)-81 B_{1}^{2}-2 B_{6}^{2} X^{2}}{3 X^{2}\left(9 B_{1}-2 B_{3}\right)}, \nonumber\\ 
 &&B_{10}=\left[18 B_{6} X\left(-12 B_{1} B_{3}+27 B_{1}^{2}+2 B_{3}^{2}\right)-36 B_{3}\left(-8 B_{1} B_{3}+18 B_{1}^{2}+B_{3}^{2}\right)\right. \nonumber\\ &&-12 B_{3} B_{6}^{2} X^{2}+4 B_{6}^{3} X^{3} ]\left[9 X^{3}\left(9 B_{1}-2B_{3}\right)^{2}\right]^{-1}.
 \end{eqnarray}

 The Class-${}^3$M-III DHOST theory
 is given by 
 \begin{eqnarray}
&&B_{2}=-\frac{9}{2} B_{1},\quad B_{3}=\frac{9}{2} B_{1}, \quad B_{4}=-\frac{3 B_{1}}{X}, \quad B_{5}=\frac{9}{X} B_{1}, \quad B_{6}=\frac{9 B_{1}}{2 X}, \nonumber\\ 
&&B_{7}=-\frac{27}{2 X} B_{1}, \quad B_{8}=\frac{9 B_{1}}{2 X^{2}}, \qquad B_{9}=-\frac{3 B_{1}}{2 X^{2}}, \qquad B_{10}=-\frac{B_{1}}{X^{3}}.
 \end{eqnarray}
 
The Class-${}^3$M-IV DHOST theory
 is given by 
 \begin{eqnarray}
B_{2}=-\frac{9}{2} B_{1}, \qquad B_{3}=\frac{9}{2} B_{1}, \quad B_{6}=-3 B_{4}-\frac{9}{2 X} B_{1}, \nonumber\\ B_{7}=-3 B_{5}+\frac{27}{2 X} B_{1}, \quad B_{9}=\frac{3 B_{1}-2 X\left(2 B_{4}+B_{5}\right)}{2 X^{2}}.
 \end{eqnarray}

The Class-${}^3$M-V DHOST theory
 is given by 
 \begin{eqnarray}
 B_{2}=B_{3}=B_{5}=B_{6}=B_{7}=B_{8}=0, \qquad B_{9}=\frac{B_{4}^{2}}{3 B_{1}}, \qquad B_{10}=\frac{B_{4}^{3}}{27 B_{1}^{2}}.
 \end{eqnarray}

The Class-${}^3$M-VI DHOST theory
 is given by 
 \begin{eqnarray}
 B_{2}=B_{3}=B_{6}=B_{7}=0.
 \end{eqnarray}
 
The Class-${}^3$M-VII DHOST theory
 is given by 
 \begin{eqnarray}
 B_{1}=B_{2}=B_{3}=B_{4}=B_{6}=0, \qquad B_{9}=-\frac{B_{5}}{X}.
 \end{eqnarray}

\subsection{The nonminimally coupled case $F_3\neq 0$}

In the nonminimally coupled case $F_3\neq  0$,
degeneracy conditions are classified into two classes \cite{BenAchour:2016fzp}:

The Class-${}^3$N-I DHOST theory ($B_1\neq 0$)
is given by 
\begin{eqnarray}
\label{3n1}
B_2&=&-3B_1,
\qquad 
B_3= 2B_1,
\qquad 
B_6=-B_4,
\qquad 
B_5
=
\frac{2(F_{3X}-3B_1)^2-2B_4 F_{3X}X}
      {3B_1X},
\nonumber\\
B_7
&=&
\frac{2B_4 F_{3X}X-2(F_{3X}-3B_1)^2}
       {3B_1 X},
\qquad
B_8
=\frac{2(3B_1+B_4X-F_{3X})\left((F_{3X}-3B_1)^2-B_4 F_{3X}X \right) }
        {9B_1^2 X^2},
\nonumber\\
B_9
&=&
\frac{2B_4(3B_1+B_4 X-F_{3X})}{3B_1X},
\qquad
B_{10}
=\frac{2B_4(3B_1+B_4 X-F_{3X} )^2 }{9B_1^2 X^2}.
\end{eqnarray}
The pure quintic order GLPV theories ($G_5\neq 0$ and $f_5\neq 0$ while $G_4=f_4=0$) satisfy the degeneracy conditions of the Class-${}^3$N-I DHOST
\cite{BenAchour:2016fzp}.

The Class-${}^3$N-II DHOST theory ($B_1= 0$) is given by 
\begin{eqnarray}
\label{3n2}
B_1=B_2=B_3=0,
\qquad
B_7=-B_5,
\qquad 
B_4=-B_6
=\frac{F_3}{X},
\qquad 
B_9= -\frac{2F_{3X}+X B_5} {X^2}.
\end{eqnarray}

\section{The full DHOST theories}
\label{app_c}

If both quadratic and cubic terms exist, 
$A_I\neq 0$ ($I=1,2,\cdots, 5$) and $B_J\neq 0$ ($J=1,2,\cdots, 10$),
they \\can be compatible only for particular combinations of Class-${}^2$N/M and Class-${}^3$N/M theories.

The compatibility of cubic DHOST theories
with Class-${}^{2}$M DHOST theories
is summarized as follows \cite{BenAchour:2016fzp}:
\begin{center}
\begin{tabular}{ c| c c c c c c c | c c}
 Class& ${}^3$M-I &${}^3$M-II & ${}^3$M-III & ${}^3$M-IV &
${}^3$M-V & ${}^3$M-VI & ${}^3$M-VII & ${}^3$N-I & ${}^3$N-II\\ 
 \hline
${}^2$M-I & \eqref{add-1} & \eqref{add-2} & \checkmark & $-$ & \eqref{add-3} &  $-$ & $-$ & $-$ & $-$ \\  
${}^2$M-II& $-$  & $-$  & \checkmark &  \checkmark & $-$ & $-$ & \eqref{add-4} &  $-$& $-$ \\ 
${}^2$M-III &$-$ & $-$ & $-$ & $-$ & \checkmark & \checkmark & \eqref{add-5} & $-$ & $-$\\  
\end{tabular}
\end{center}
where we have defined
\begin{eqnarray}
\label{add-1}
B_4
&=&
\frac{-6A_1B_1+4A_2B_2+A_3 X(9B_1+2B_2)}
      {2X(A_1+3A_2)},
\\
\label{add-2}
B_6
&=&
\frac{3(6A_1B_1+4A_2B_3+A_3 X(2B_3-9B_1))}
      {4X(A_1+3A_2)},
\\
\label{add-3}
B_4
&=&
-\frac{3B_1 (2A_1-3A_3 X)}
        {2X(A_1+3A_2)},
\\
\label{add-4}
B_7
&=&
-3B_5,
\\
\label{add-5}
B_7&=&0.
\end{eqnarray}

The compatibility of cubic DHOST theories
with Class-${}^{2}$N DHOST theories
is summarized as follows \cite{BenAchour:2016fzp}:
\begin{center}
\begin{tabular}{ c| c c c c c c c | c c}
 Class& ${}^3$M-I &${}^3$M-II & ${}^3$M-III & ${}^3$M-IV &
${}^3$M-V & ${}^3$M-VI & ${}^3$M-VII & ${}^3$N-I & ${}^3$N-II\\ 
 \hline
${}^2$N-I & \eqref{add1} \& \eqref{add3} & \eqref{add1} \& \eqref{add6} & \eqref{add1} &  $-$& \eqref{add1} \& \eqref{add4} & $-$ & $-$& \eqref{add8} & $-$ \\  
${}^2$N-II & $-$ & $-$ & $-$ & $-$& $-$ & $-$ & 
\eqref{add7}
& $-$ &  \checkmark\\ 
${}^2$N-III & \eqref{add3} & \eqref{add6} & \checkmark & $-$ & \eqref{add4} & $-$ & $-$& $-$ & $-$\\  
${}^2$N-IV & \eqref{add2} \& \eqref{add3} & \eqref{add2} \& \eqref{add6} & \eqref{add2} &  $-$& \eqref{add5} & $-$ & $-$  & $-$ & $-$ 
\end{tabular}
\end{center}
where we have defined
\begin{eqnarray}
\label{add1}
A_3
&=&
-\frac{8A_1F_{2X}}{F_2}
+\frac{6A_1+4F_{2X}}{X}
-\frac{4F_2}{X^2},
\\
\label{add2}
A_3
&=&
\frac{12A_2 F_{2X}}{F_2}
-\frac{8(A_2-F_{2X})}{X}
-\frac{6F_2}{X^2},
\\
\label{add3}
B_4
&=&\frac{2F_{2X} (9B_1+2B_2)}
        {F_2}
-\frac{2(6B_1+B_2)}{X},
\\
\label{add4}
B_4
&=&
6B_1
\left(
\frac{3F_{2X}}{F_2}
-\frac{2}{X}
\right),
\\
\label{add5}
B_4
&=&
\frac{3B_1}{2X(A_2 X+F_2)}
 \left(
X(A_3 X+4F_{2X})
-2F_{2}
\right),
\\
\label{add6}
B_6
&=&\frac{3}{F_2X}
\left(
6B_1F_2-9B_1F_{2X}X-B_3F_2+2B_3F_{2X}+2B_3 F_{2X}X
\right),
\\
\label{add7}
B_7&=&-B_5,\\
\label{add8}
B_4
&=&
\frac{-A_1F_{3X} X-6B_1F_2+6B_1 F_{2X}X+2F_2F_{3X}}
      {F_2X},\\
A_3
&=&
\frac{2}{3B_1 F_2 X^2}
\left[
 B_1 (9A_1F_2X-12 A_1F_{2X} X^2+6F_2F_{2X}X-6F_{2X}^2)
+2F_{3X} (F_2-A_1 X)^2
\right].
\end{eqnarray}

\section{The coefficients in general couplings}

\subsection{The pure cubic DHOST theories}

In the case of the pure cubic DHOST theories,
the coefficients in the Taylor series form \eqref{taylor-c} satisfy
the following relations:

Case-1-cubic satisfies 
\begin{eqnarray}
\label{coe-1c}
&&
f_{00}=f_{01}=f_{11}=0,
\qquad
b_{20}-9b_{10}
=
b_{30}+18 b_{10}
=0,
\nonumber\\
&&
b_{31}=
\frac{3}{10X_0}
\left(
-2b_{20}+b_{30}
-3X_0
(2b_{11}+2b_{21}+b_{40}+b_{60})
\right),
\end{eqnarray}
with $X_0=-q^2$.

Case-2-cubic satisfies
\begin{eqnarray}
\label{coe-2c}
f_{00}=f_{01}=f_{11}=0,
\qquad
b_{10}=b_{20}=b_{30}=b_{31}=b_{60}=0,
\qquad
2b_{11} + b_{21}=0,
\qquad 
b_{21}+ b_{40}=0.
\end{eqnarray}

Case-1-$\Lambda$-cubic satisfies 
\begin{eqnarray}
\label{coe-1lc}
&&
f_{00}=-\Lambda M_p^2,
\qquad
f_{01}=0,
\qquad
b_{30}=b_{20}=b_{30}=b_{31}=0,
\nonumber\\
&&
 b_{40}+b_{60}
+2\Lambda (b_{11}+b_{21})
=0,
\quad
f_{11}
-\Lambda
\left(
-X_0(b_{40}-4b_{11}-2b_{21})
+f_{31}
\right)
=0,
\end{eqnarray}
with $X_0=-q^2$.

Case-2-$\Lambda$-cubic satisfies 
\begin{eqnarray}
\label{coe-2lc}
&&f_{00}=-\Lambda M_p^2,
\qquad
f_{01}=0,
\qquad
f_{11}+\Lambda (2X_0 b_{11}-f_{31})=0,
\nonumber\\
&&b_{30}=b_{20}=b_{30}=b_{31}=b_{60}=0,
\qquad
b_{21}+2b_{11}=b_{40}-2b_{11}=0.
\end{eqnarray}

\subsection{The full DHOST theories}

In the case of the pure cubic DHOST theories,
the coefficients in the Taylor series form \eqref{taylor-f} satisfy
the following relations:

Case-1-full satisfies 
\begin{eqnarray}
\label{coe-1f}
&&
f_{00}=f_{01}=f_{11}=0,
\qquad
  a_{10}+a_{20}
= a_{11}+a_{21}
=0,
\qquad
b_{20}-9b_{10}
=
b_{30}+18 b_{10}
=0,
\nonumber\\
&&
b_{31}=
\frac{3}{10X_0}
\left(
-2b_{20}+b_{30}
-3X_0
(2b_{11}+2b_{21}+b_{40}+b_{60})
\right),
\end{eqnarray}
with $X_0=-q^2$.

Case-2-full satisfies
\begin{eqnarray}
\label{coe-2f}
&&
f_{00}=f_{01}=f_{11}=0,
\qquad
a_{10}
=a_{20}
=a_{11}+a_{21}
=a_{30}+2a_{11}
=0,
\nonumber\\
&&
b_{10}=b_{20}=b_{30}=b_{31}=b_{60}=0,
\nonumber\\
&&
2b_{11} + b_{21}=0,
\qquad 
b_{21}+ b_{40}=0.
\end{eqnarray}

Case-1-$\Lambda$-full satisfies 
\begin{eqnarray}
\label{coe-1lf}
&&
f_{00}+2\Lambda f_{20}-2X_0 \Lambda a_{10}=0,
\qquad
f_{01}+ \Lambda 
\left(
4f_{21}
-a_{10}
+2X_0 a_{11}
+\frac{3}{2}X_0 a_{30}
\right)
=0,
\nonumber\\
&&
f_{11}
-\Lambda
\left(
-X_0 b_{40}
+4X_0 b_{11}
+2X_0 b_{21}
+f_{31}
\right)
=0,
\qquad
 a_{10}+a_{20}
=a_{11}+a_{21}
=0,
\nonumber\\
&&
b_{10}=b_{20}=b_{30}=b_{31}=0,
\qquad
 b_{40}+b_{60}
+2\Lambda (b_{11}+b_{21})
=0,
\end{eqnarray}
with $X_0=-q^2$.

Case-2-$\Lambda$-full satisfies 
\begin{eqnarray}
\label{coe-2lf}
&&f_{00}+2\Lambda f_{20}=0,
\qquad
f_{01} +\Lambda (4f_{21}-X_0 a_{11})=0,
\qquad
f_{11}+\Lambda (2X_0 b_{11}-f_{31})=0,
\nonumber\\
&&
a_{10}
=a_{20}
=a_{11}+a_{21}
=a_{30}+2a_{11}
=0,\nonumber\\
&&
b_{10}=b_{20}=b_{30}=b_{31}=b_{60}=0,
\qquad
b_{21}+2b_{11}=b_{40}-2b_{11}=0.
\end{eqnarray}

\bibliography{ref-B}

\end{document}